\documentclass[journal,a4paper]{IEEEtran}
\def\IEEEkeywordsname{Keywords}
\ifCLASSINFOpdf
  \usepackage[pdftex]{graphicx}
\else
\fi
\usepackage[cmex10]{amsmath}
\usepackage{array}
\usepackage[caption=false,font=footnotesize]{subfig}
\usepackage{fixltx2e}
\usepackage{stfloats}

\usepackage{multirow}

\usepackage{tikz}
\usepackage{pgfplots}

\begin{document}
%
% paper title
% can use linebreaks \\ within to get better formatting as desired
%\title{Occlusion Handling in Depth Estimation from Multiview Video}
\title{Depth Estimation using Modified Cost Function for Occlusion Handling}
%
%
% author names and IEEE memberships
% note positions of commas and nonbreaking spaces ( ~ ) LaTeX will not break
% a structure at a ~ so this keeps an author's name from being broken across
% two lines.
% use \thanks{} to gain access to the first footnote area
% a separate \thanks must be used for each paragraph as LaTeX2e's \thanks
% was not built to handle multiple paragraphs
%

\author{Krzysztof~Wegner, Olgierd~Stankiewicz and Marek~Domanski% <-this % stops a space
\thanks{Krzysztof~Wegner, Olgierd~Stankiewicz and Marek~Domanski are with the Chair of Multimedia Telecommunications and Microelectronics, Poznan University of Technology, Poznan, Poland (e-mail: kwegner@multimedia.edu.pl)}% <-this % stops a space        
%\thanks{M. Shell is with the Department
%of Electrical and Computer Engineering, Georgia Institute of Technology, Atlanta,
%GA, 30332 USA e-mail: (see http://www.michaelshell.org/contact.html).}% <-this % stops a space
%\thanks{J. Doe and J. Doe are with Anonymous University.}% <-this % stops a space
}

% note the % following the last \IEEEmembership and also \thanks -
% these prevent an unwanted space from occurring between the last author name
% and the end of the author line. i.e., if you had this:
%
% \author{....lastname \thanks{...} \thanks{...} }
%                     ^------------^------------^----Do not want these spaces!
%
% a space would be appended to the last name and could cause every name on that
% line to be shifted left slightly. This is one of those \L{}aTeX things". For
% instance, "\textbf{A} \textbf{B}" will typeset as \k{A} B" not \k{A}B". To get
% \k{A}B" then you have to do: "\textbf{A}\textbf{B}"
% \thanks is no different in this regard, so shield the last } of each \thanks
% that ends a line with a % and do not let a space in before the next \thanks.
% Spaces after \IEEEmembership other than the last one are OK (and needed) as
% you are supposed to have spaces between the names. For what it is worth,
% this is a minor point as most people would not even notice if the said evil
% space somehow managed to creep in.

% The paper headers
%\markboth{Journal of \LaTeX\ Class Files,~Vol.~6, No.~1, January~2007}%
%{Shell \MakeLowercase{\textit{et al.}}: Bare Demo of IEEEtran.cls for Journals}%
%
\markboth{}{}
%
% The only time the second header will appear is for the odd numbered pages
% after the title page when using the twoside option.
%
% *** Note that you probably will NOT want to include the author's ***
% *** name in the headers of peer review papers.                   ***
% You can use \ifCLASSOPTIONpeerreview for conditional compilation here if
% you desire.

% If you want to put a publisher's ID mark on the page you can do it like
% this:
%\IEEEpubid{0000--0000/00\$00.00~\copyright~2007 IEEE}
% Remember, if you use this you must call \IEEEpubidadjcol in the second
% column for its text to clear the IEEEpubid mark.

% use for special paper notices
%\IEEEspecialpapernotice{(Invited Paper)}

% make the title area
%
\pagestyle{empty}%
\maketitle%
\thispagestyle{empty}%

\begin{abstract}
%\boldmath
The paper presents a novel approach to occlusion handling problem in depth estimation using three views. A solution based on modification of similarity cost function is proposed. During the depth estimation via optimization algorithms like Graph Cut similarity metric is constantly updated so that only non-occluded fragments in side views are considered. At each iteration of the algorithm non-occluded fragments are detected based on side view virtual depth maps synthesized from the best currently estimated depth map of the center view. Then similarity metric is updated for correspondence search only in non-occluded regions of the side views. The experimental results, conducted on well-known 3D video test sequences, have proved that the depth maps estimated with the proposed approach provide about 1.25 dB virtual view quality improvement in comparison to the virtual view synthesized based on depth maps generated by the state-of-the-art MPEG Depth Estimation Reference Software. 
\end{abstract}
% IEEEtran.cls defaults to using nonbold math in the Abstract.
% This preserves the distinction between vectors and scalars. However,
% if the journal you are submitting to favors bold math in the abstract,
% then you can use LaTeX's standard command \boldmath at the very start
% of the abstract to achieve this. Many IEEE journals frown on math
% in the abstract anyway.

% Note that keywords are not normally used for peerreview papers.

% For peer review papers, you can put extra information on the cover
% page as needed:
% \ifCLASSOPTIONpeerreview
% \begin{center} \bfseries EDICS Category: 3-BBND \end{center}
% \fi
%
% For peerreview papers, this IEEEtran command inserts a page break and
% creates the second title. It will be ignored for other modes.
\IEEEpeerreviewmaketitle

\newcounter{MYtempeqncnt}

\section{Introduction}
\IEEEPARstart{3D}{} video systems have recently gained a lot of attention. Many new 3D video systems have been developed. Among them super multiview television and free viewpoint television can be examples of such novel 3D systems. In the free viewpoint television a user is able to freely choose a position of a virtual camera. The requested view of a scene is generated from dynamic 3D representation of the scene. 

The most commonly used 3D representation is a MultiVideo and Depth (MVD) \cite{bib:mvd} composed of multiple videos acquired by the set of cameras and accompanied depth maps for each of the views. Based on transmitted videos and depth data any view can be easily generated by employing depth-image-base rendering (DIBR) \cite{bib:dibr}.

Recently 3D extension of such standards as AVC \cite{bib:mvc,bib:3davc} and HEVC \cite{bib:3dhevc} that allows for efficient transmission of dynamic 3D scene representation in MVD format has been finalized. 

Depth information in such systems can be acquired either directly by depth cameras \cite{bib:tof}, or indirectly by algorithmic depth estimation from recorded videos \cite{bib:naszftv}. Commonly depth information is obtained by the conversion from disparity information \cite{bib:hartley}. Although in computer vision, disparity $d$ is often treated as synonymous with depth (distance $z$), essentially those terms are the inverse of each other. 

\begin{equation}
z \sim \frac{1}{d}
\end{equation}

Disparity is a displacement vector between corresponding fragments (pixels, blocks) of two images of the same scene taken from different viewpoints. Those two corresponding fragments represent the same fragment of an observed scene but seen from two different viewpoints. 

\begin{figure}[!h]
\centering
\includegraphics[width=4cm]{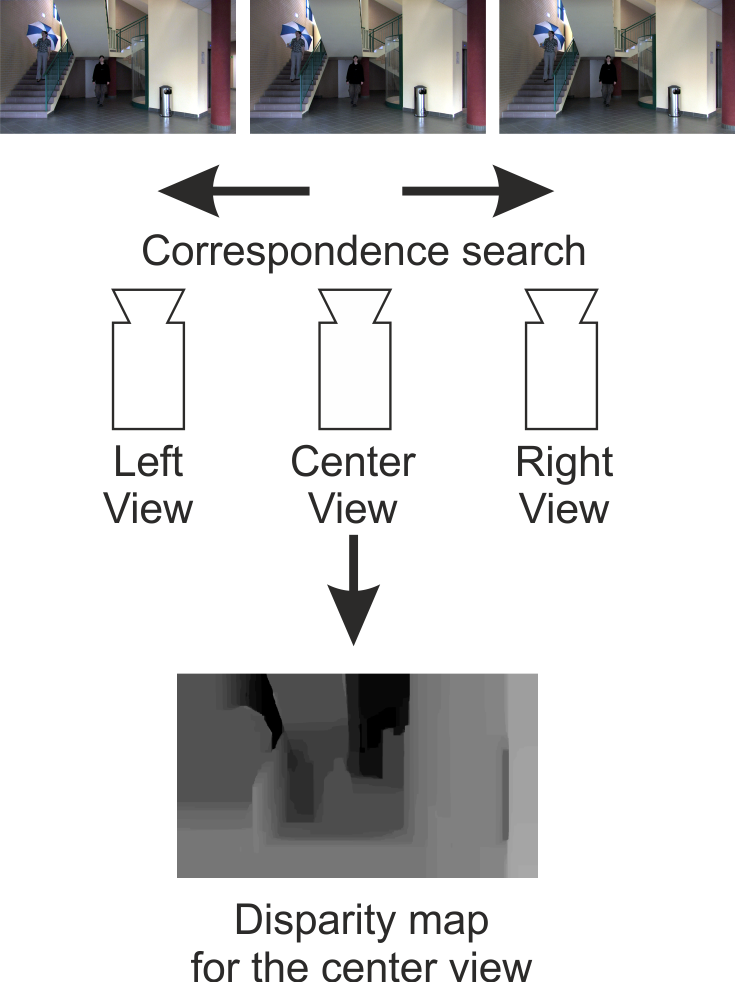}
\caption{Three-view disparity estimation.}
\label{fig:3viewde}
\end{figure}

Stereo correspondence search is an active research topic in computer vision, and one of the basic method of obtaining disparity information. There are many stereo disparity estimation methods known. Comprehensive study of stereo disparity estimation methods can be found in \cite{bib:steromatching}, and on the Middlebury webpage \cite{bib:middleweb} containing up-to-date benchmark of stereo disparity estimation methods. In the scope of development of multiview systems stereo correspondence search was extended to multiview correspondence search \cite{bib:muliviewmatching1,bib:muliviewmatching2, bib:multiviewmatching}. 

For the sake of simplicity and accuracy, many algorithms assume that images are taken by a rectified set of cameras \cite{bib:rectify1,bib:rectify2}. Consecutively, corresponding fragments of a given image can be found on the same horizontal line in the remaining images. 

Some algorithms use three views (left, central and right)  \cite{bib:dersold,bib:de3view-tanimoto,bib:de3view-ho, bib:os} as inputs and produce disparity map or depth map for the central view (Fig.~\ref{fig:3viewde}). Often when it is not important which of left or right view is referred to, a name "side view" is used instead.

During disparity estimation, for a given fragment of the central view, the algorithm searches for the corresponding fragment in the side views that represent the same fragment/portion of the scene. 

The correspondence search is done on the basis of Similarity Metric which expresses how probable it is that a certain fragment of one image is the corresponding fragment of the second image. Although the metric used is often called similarity, it actually expresses dissimilarity between fragments. There are many Similarity Metrics known from literature: Sum of Absolute Difference (SAD) Sum of Squared Difference (SSD), Rank, Census, Cross Correlation and other \cite{bib:depth_similarity, bib:depth_similarity2}.

The correspondence search is often defined as an optimization problem in which for every fragment of the central image the best (the most similar) fragment of the side views is selected. This optimization problem maybe expressed in terms of energy function using Markov Random Field (MRF) and optimized via one of the optimization algorithms such as Belief Propagation \cite{bib:bp1,bib:bp2}, Dynamic Programming \cite{bib:dp}, or Graph Cut \cite{bib:gc}. 

Since input videos are captured by multiple cameras with different positions, some parts of the observed scene can be occluded, and thus not visible, within some of the views. Disparity estimation for those fragments of a scene is challenging and requires special care. 
If the the algorithm do not properly taking into account, possible occlusions within the scene, estimated disparity can be wrong. Estimated disparity can indicated not truly corresponding fragments. 
%If the correspondence search is not equipped with proper occlusion handling, obtained results can be inadequate, and lead to wrong estimated disparity.

In this paper a novel approach to occlusion handling designed to work in three-view disparity estimation algorithms is proposed. 
  
\section{Occlusion problem in disparity estimation}

\begin{figure}[!b]
\centering
\includegraphics[width=8cm]{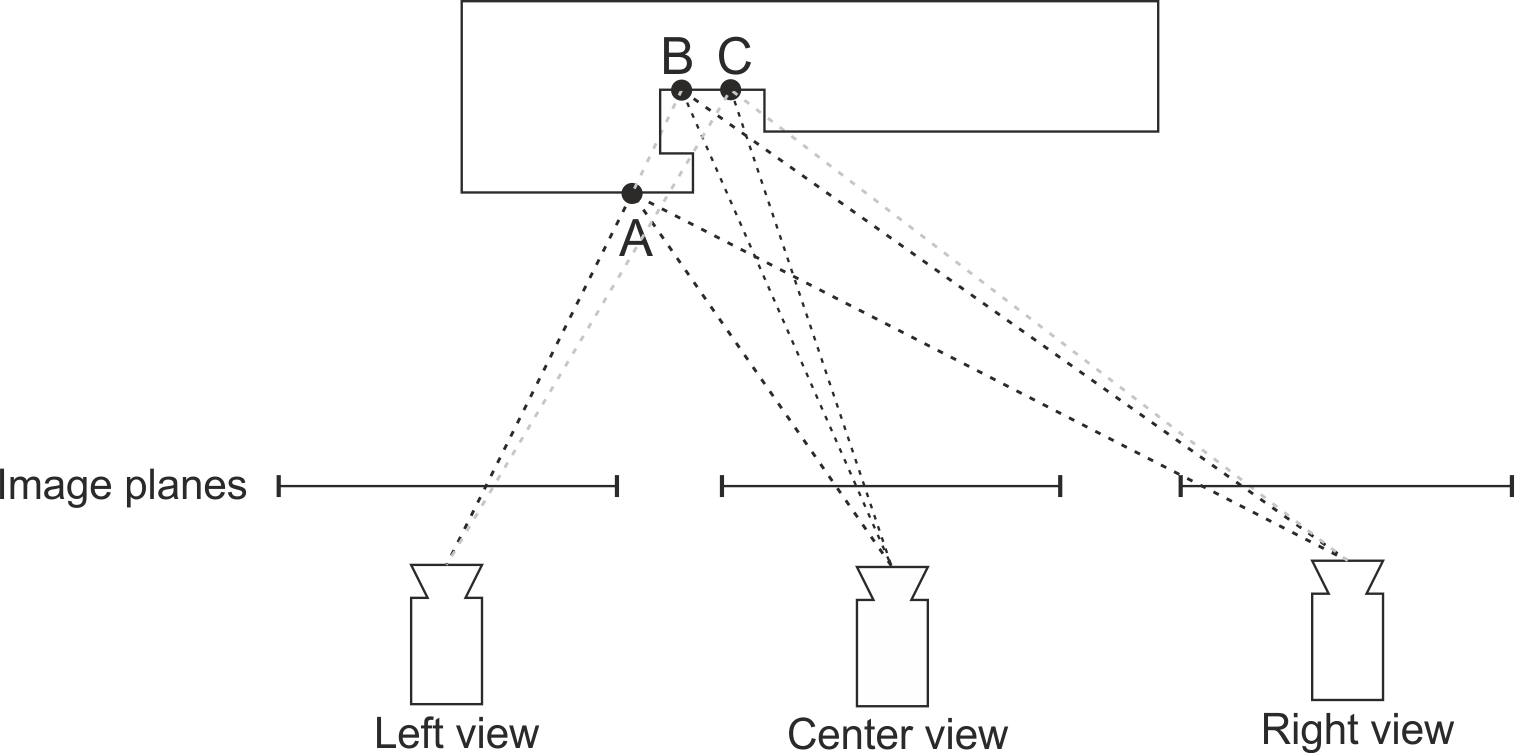}
\caption{Occlusion in three-view disparity estimation problem.}
\label{fig:occ}
\end{figure}

\begin{figure*}[!t]
% ensure that we have normalsize text
\normalsize
% Store the current equation number.
\setcounter{MYtempeqncnt}{\value{equation}}%\newcounter{NameOfTheNewCounter}
% Set the equation number to one less than the one
% desired for the first equation here.
% The value here will have to changed if equations
% are added or removed prior to the place these
% equations are referenced in the main text.
\setcounter{equation}{4}
\small
%
%Cost(x,y,t) = \left\{ \begin{matrix} \frac{NotOcc_L(x,y,t) \cdot SimMetric(I_{C}(x,y),I_{L}(x+t,y)) + NotOcc_R(x,y,t) \cdot SimMetric(I_{C}(x,y),I_{R}(x-t,y))}{NotOcc_L(x,y,t)+NotOcc_R(x,y,t)} & for & NotOcc_L(x,y,t)+NotOcc_R(x,y,t) \neq 0\\ const & for & NotOcc_L(x,y,t)+NotOcc_R(x,y,t)=0 \end{matrix} \right.
\begin{equation}
\label{eqn:novelcost}
Cost(x,y,t) = \frac{NotOcc_{L}(x,y,t) \cdot Sim(I_{C}(x,y),I_{L}(x+t,y) + NotOcc_{R}(x,y,t) \cdot Sim(I_{C}(x,y),I_{R}(x-t,y)))}{NotOcc_{L}(x,y,t) + NotOcc_{R}(x,y,t)}
\end{equation}
\normalsize
% Restore the current equation number.
\setcounter{equation}{\value{MYtempeqncnt}}
% IEEE uses as a separator
\hrulefill
% The spacer can be tweaked to stop underfull vboxes.
\vspace*{4pt}
\end{figure*}

Given three images, center $I_{C}$, left $I_{L}$ and right $I_{R}$ of the same size, we search for such a displacement $t$ for every pixel P of center view (at coordinates $(x,y)$) that minimize cost function expressing a similarity between pixel P (or small fragment around the pixel P like block) and a corespondent pixel P' (small fragment around pixel P') displaced by $t$ in side views (at positions $(x+t,y)$ in left and $(x-t,y)$ in right view). Such displacement is then a disparity of a given pixel P of a center view.

\begin{equation}
d_{Center}(x,y) = \min_{t}{Cost(x,y,t)};
\label{eqn:cost}
\end{equation}

In disparity estimation based on / from three views (see Fig.~\ref{fig:occ}), a given point of the scene visible from center view can be visible from both of the side views (point A), or only from one of the side views (left or right, point B), or from neither of them (point C). 

If the given fragment of the scene visible from center view is not visible from one or both of the side views, we say that a given fragment of the scene is occluded in side view (is not visible from that particular side view). 

The simplest method for detecting occluded fragments is cross-checking \cite{bib:crosscheck}. Cross-checking tests the consistency of estimated disparity value for pixels from center view with those estimated for pixels in left and right views. If the disparity value estimated in each view is different for a correspondent triple of pixels from center, left and right views given pixels are assumed to be occluded. Next, the disparity value for occluded pixels are extrapolated from neighboring pixels that are not occluded. In order to perform cross-checking, disparity maps for all of the three views are required. Estimation of three disparity maps is not always possible. Even if the  estimation of three disparity maps instead of one is possible it is resource and time consuming. 

%Many be explain whay the consistency menas.

Occlusion handling is performed by adding/putting additional constraints, such as ordering constraint or uniqueness constraint to objective function of optimization procedures like Graph Cut (GC), Dynamic Programming (DP) or Belief Propagation (BP) used to estimate the disparity map. 

The ordering constraint \cite{bib:ordering} imposes the same order of corresponding pixels in all views. If a pixel A is on the left of pixel B in the center view, in the side view pixel A' that is a corresponding pixel of pixel A must be as well on the left of pixel B', a corresponding pixel of pixel B. In real scenes the ordering constraint can be violated in the case of big perspective change or in case of thin objects. In such cases ordering constraint can introduce errors in estimated disparity maps. 

The uniqueness constraint \cite{bib:uniqness,bib:uniqness2} imposes the one-to-one correspondence between pixel in center and side views. If a given pixel A of the central view is assigned to a corresponding pixel B in the side view, no other pixel of central view can be assigned to a correspondence with pixel B in side view. This way unique pixel to pixel correspondence is forced across all of the views. 

There are many disparity estimation algorithms known, that handle occlusion in efficient way \cite{bib:occ1,bib:occ2,bib:uniqness2}. The main drawback of all of those algorithms are additional constraints (terms) imposed in optimization procedures with increased complexity and thus execution time of the disparity estimation. 

Another approach to occlusion handling is to change cost term (eq.~\ref{eqn:cost}) composed of similarity metric in optimization algorithms. As we search for corresponding fragment of a central view in both side views simultaneously, there are many ways of defining a $Cost(x,y,t)$ function. 

Commonly \cite{bib:de3view-tanimoto,bib:de3view-ho} it is the sum of similarity metrics between a fragment in the center view and corresponding fragments in left and right views.

\begin{equation}
\begin{matrix}
Cost(x,y,t) &= Similarity(I_{C}(x,y),I_{L}(x+t,y))\\
 &+ Similarity(I_{C}(x,y),I_{R}(x-t,y))
\end{matrix}
\end{equation}

Because of the occlusions Tanimoto \cite{bib:dersold} proposed to pick just the most similar fragment from either left or right view. The intuition is that the occluded fragment of the images will lead to less similar fragment, thus the minimum of similarity metrics from left and right view is used.

\begin{equation}
\begin{matrix}
Cost(x,y,t) &= min(&Similarity(I_{C}(x,y),I_{L}(x+t,y)), \\
            &      &Similarity(I_{C}(x,y),I_{R}(x-t,y)))
\end{matrix}
\end{equation}

In this paper we propose yet another way to define cost function which takes into account an occlusion possible within the scene.

\section{Proposed occlusion handling}

\begin{figure}[!b]
\centering
\includegraphics[width=8cm]{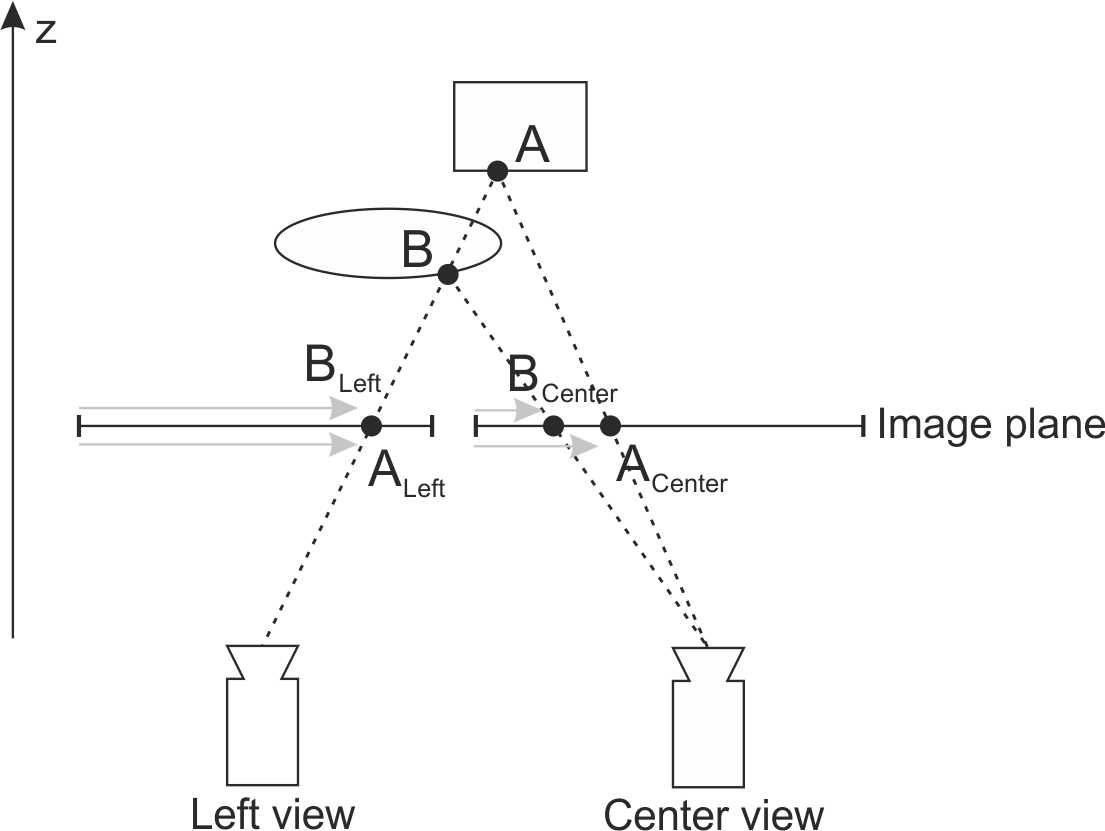}
\caption{Occlusion problem in correspondence search.}
\label{fig:occ_detection}
\end{figure}

As it was said before, a given fragment of a scene visible from center view can be occluded in one or both side views (left or/and right) (Fig.~\ref{fig:occ}). In such a case, searching for a correspondence of a given pixel of center view in this particular side view (left or right) is pointless, as the given fragment of the scene is not visible from that particular side view. Considering the correspondence with an occluded fragment of an image could cause errors in estimated disparity.

Therefore the correspondence search should be performed only in side views in which a considered fragment of a center view is not occluded. The cost function should be constructed in such a way that it considers only similarity metrics from not occluded views. If a given fragment is visible in both views, then the cost function should be an average of both similarity metrics, in order to reduce the influence of noise, which is present in all views. We propose to define the cost function in a way that it considers only similarity metrics of fragments from a not occluded view (either left or right) (eq. \ref{eqn:novelcost}) where  $NotOcc_{L}(x,y,t)$, $NotOcc_{R}(x,y,t)$ expresses whether a given pixel of a center view is not occluded in left and right views respectively. 
Depending on the existence of occlusion in the views, the sum $NotOcc_{L}(x,y,t) + NotOcc_{R}(x,y,t)$ in the denominator of eq. \ref{eqn:novelcost} can be 2 if a pixel in not occluded in both views, 1 if it is occluded in one of the side views (either left or right), and 0 if it is occluded in both side views. If a given pixel is occluded in both side views the equation \ref{eqn:novelcost} loses its meaning, thus in such a case constant penalty value is used as a cost value.
\setcounter{equation}{5}
\begin{equation}
Cost(x,y,t) = const
\end{equation}

But why a given fragment (object A) of a scene is not visible in a side view? Because in a side view that fragment is occluded by some other part of the scene (object B). Object B blocks light rays from object A, so in side view closer object B is visible instead the farther object A.

Consider the example on Fig.~\ref{fig:occ_detection} where two points A and B are observed by two cameras (left and center). Point B is closer to the cameras and point A is farther. Point B is visible in both views (left and center) at pixel position $B_{Left}$ and $B_{Center}$ respectively. But due to the occlusion , point A is visible only in center view at pixel position $A_{Center}$. If there would be no point B, point A should/would be visible in left view at pixel position $A_{left}$. The disparity of point B in left view is the difference of the pixel position $B_{Left}$ and $B_{Center}$ and disparity of point A in left view would be (if the point was/would be visible) the difference of pixel position $A_{Left}$ and $A_{Center}$.

\begin{equation}
d_{Left}(B_{Left})=B_{Left}-B_{Center}
\end{equation}

\begin{equation}
d_{Left}(A_{Left})=A_{Left}-A_{Center}
\end{equation}

The distance to the camera $z$ is reciprocal to disparity. So a fragment of an image representing a closer object (point B) has bigger disparity than the fragment representing farther object (point A).

\begin{equation}
z_{Left}(A_{Left}) > z_{Left}(B_{Left}) <=> d_{Left}(A_{Left}) < d_{Left}(B_{Left})
\end{equation}

For a given pixel $A_{Center}$ of center view at coordinates $(x,y)$ and considered displacement $t$, corresponding pixel $A_{Left}$ in left view should be at coordinates $(x+t,y)$. So, if we want to check whether a fragment A of a scene is occluded in left view we have to check the disparity (distance) assigned to the considered corresponding pixel $A_{Left}$ in left view. If a disparity $d_{Left}(x+t,y)$ assigned already to considered corresponding pixel $A_{Left}$ is bigger than the considered displacement t then probably a pixel $A_{left}$ is not a fragment of the same object A but rather some other closer object B that occludes object A in the left view.

Based on such a consideration we can create a function assessing whether for a pixel at coordinates $(x,y)$ and displacement $t$, corresponding pixel is/can be/will be occluded or not in left and right views.

\begin{equation}
NotOcc_{L}(x,y,t) = \left\{ \begin{matrix} 
1 & for & t \geq d_{Left}(x+t,y)\\ 
0 & for & t < d_{Left}(x+t,y)
\end{matrix} \right.
\end{equation}

\begin{equation}
NotOcc_{R}(x,y,t) = \left\{ \begin{matrix} 
1 & for & t \geq d_{Right}(x-t,y)\\ 
0 & for & t < d_{Right}(x-t,y)
\end{matrix} \right.
\end{equation}

$NotOcc(x,y,t)$ equal 1 means that the corresponding pixel in side view at a given displacement is probably not occluded.

\section{Application of Proposed Idea}

The proposed idea is general - it does not impose any particular source of disparity maps $d_{Left}$ for left and $d_{Right}$ for right view. But in general, disparity maps for left and right views are unknown before estimating the disparity for central view.

Commonly, disparity maps are estimated iteratively with the use of such algorithms like Belief Propagation or Graph Cut. In such algorithms, at each iteration of the estimation, algorithm maintains up-to-date / best already estimated disparity map for center view. This disparity map is further refined in the next iteration of the algorithm.

For our occlusion detection we propose to use disparity maps of side views created based on the disparity map of center view through Depth-Image-Based Rendering (DIBR). After each iteration of a disparity estimation algorithm, we create disparity maps of side views ($d_{Left}$ and $d_{Right}$) from the best already estimated disparity map of a center view. This way if the estimation algorithm used assigned already some disparity $d_{Center}(B)$ to some pixel $B_{Center}$, then pixel $A_{Center}$ cannot have such a disparity that the corresponding pixel $A_{Left}$ (Fig.~\ref{fig:occ_detection}) is at the same position as corresponding pixel $B_{Left}$ of pixel $B_{Center}$. In other words fragment B of a scene represented by pixel $B_{Center}$ in center view should occlude a fragment A of a scene (represented by pixel $A_{Center}$ in center view) seen from left view.

%\begin{figure}[!t]
%\begin{figure}[!h]
%\centering
%\includegraphics[width=8cm]{fig/image2}
%\caption{The correspondence of given pixel in the center view to the pixels in the neighboring views and proposed deduction of the occlusion}
%\label{fig:occ_detection}
%\end{figure}

\section{Experiments}

\begin{table}[!t]
\caption{Positions of views used for evaluation of quality of estimated disparity maps.}
\label{table:video-views}
\centering
\begin{tabular}{|l|c|c|c|}
\hline
\textbf{Sequence Name} & \textbf{View A} & \textbf{View B} & \textbf{View V} \\ \hline
Poznan Street  & 3 & 5 & 4 \\ \hline
Poznan Hall 2  & 5 & 7 & 6 \\ \hline
Poznan CarPark & 3 & 5 & 4 \\ \hline
Book Arrival   & 7 & 9 & 8 \\ \hline
\end{tabular}
\end{table}

\begin{figure}[!t]
\centering
\includegraphics[width=8cm]{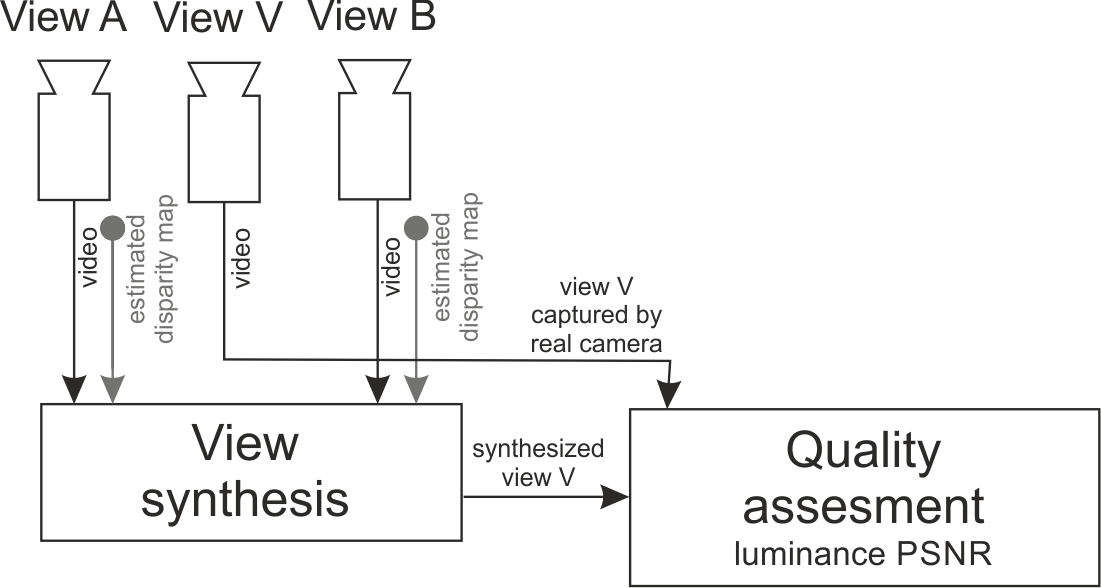}
\caption{Disparity map quality evaluation methodology.}
\label{fig:evalation}
\end{figure}

We have implemented our idea in Depth Estimation Reference Software (DERS) \cite{bib:ders} version 5.0 developed by Moving Picture Experts Group (MPEG) of International Standardization Organization (ISO) during works on 3D video compression standardization. DERS is the state-of-the-art disparity estimation technique, designed with 3D video application in mind. It uses Graph Cut as the optimization algorithm along with many other techniques that improve or/and speed up disparity estimation from three input videos. 

\begin{figure}[!h]
\subfloat[Poznan Street \cite{bib:poznan_sequences}]{\includegraphics[width=4cm]{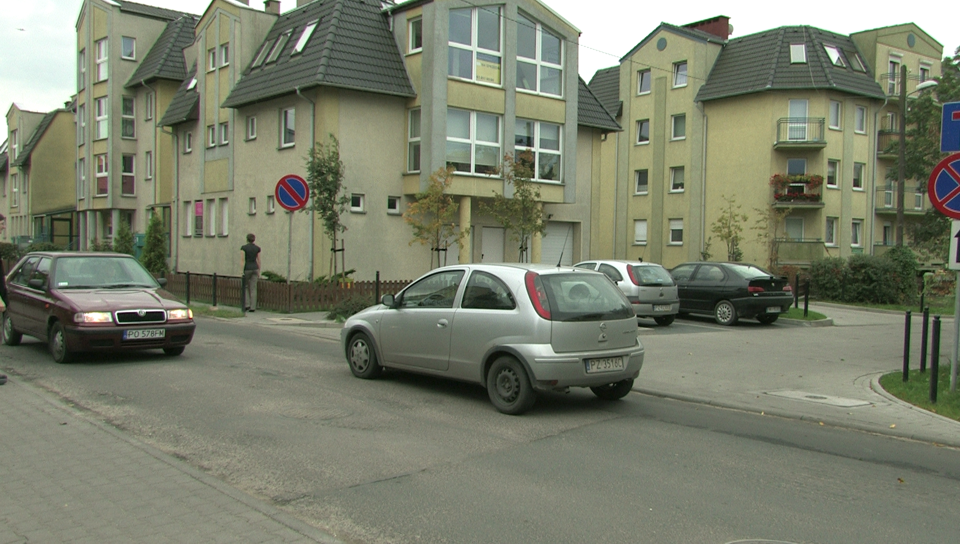}%
\label{fig_first_case2}}
\hfil
\subfloat[Poznan CarPark \cite{bib:poznan_sequences}]{\includegraphics[width=4cm]{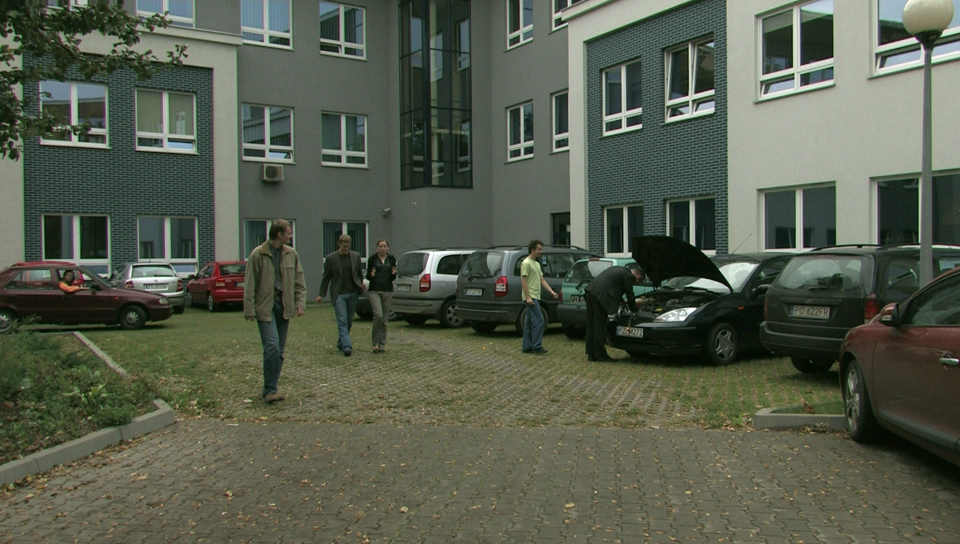}%
\label{fig_second_case2}}
\\
\subfloat[Poznan Hall2 \cite{bib:poznan_sequences}]{\includegraphics[width=4cm]{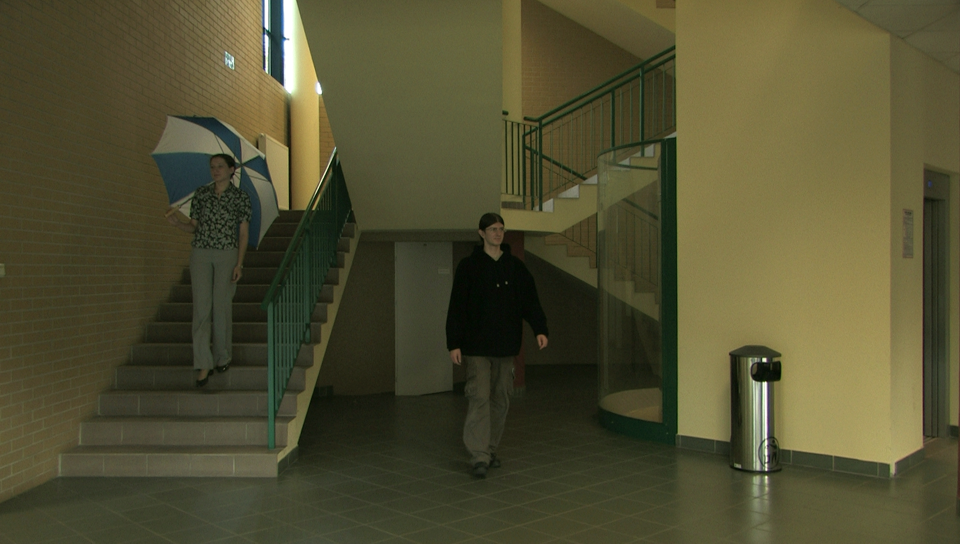}%
\label{fig_third_case2}}
\hfil
\subfloat[Book Arrival \cite{bib:hhi_sequences}]{\includegraphics[width=4cm]{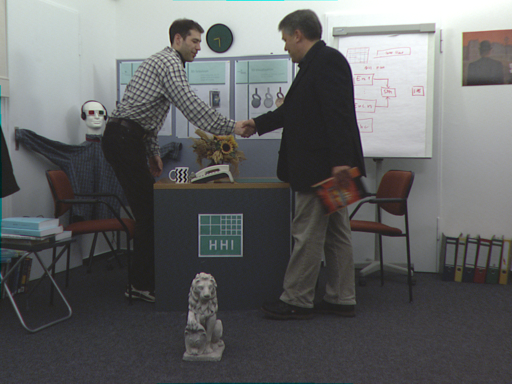}%
\label{fig_fourth_case2}}
\caption{Exemplary frames from multiview test sequences used in experiments}
\label{fig:test_sequences}
\end{figure}

Proposed approach was tested on four 3D video test sequences recommended by the MPEG committee (Fig.~\ref{fig:test_sequences}) namely: Poznan Street, Poznan CarPark, Poznan Hall 2 \cite{bib:poznan_sequences}, Book Arrival \cite{bib:hhi_sequences}. 

In applications such as Free View Television, disparity maps are used mainly for the purpose of view synthesis. Therefore, we have evaluated our proposed method indirectly, by assessing the quality of the synthesized views. 

\begin{figure}[!h]
\subfloat[Tsukuba]{\includegraphics[width=4cm]{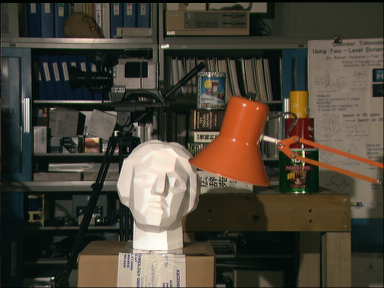}%
\label{fig_first_case}}
\hfil
\subfloat[Venus]{\includegraphics[width=4cm]{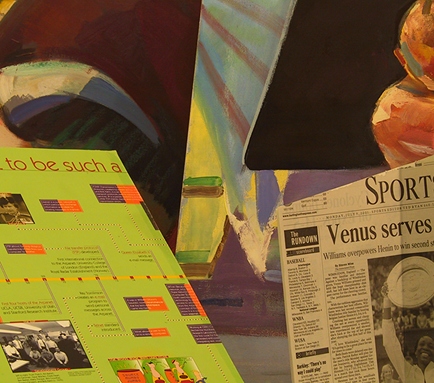}%
\label{fig_second_case}}
\\
\subfloat[Teddy]{\includegraphics[width=4cm]{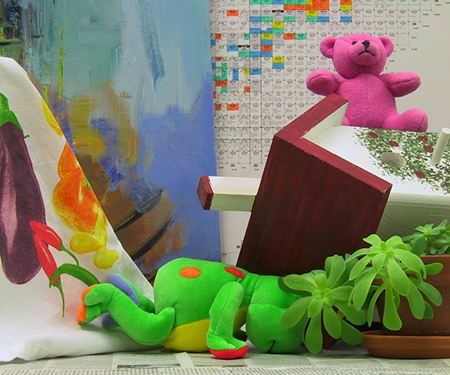}%
\label{fig_third_case}}
\hfil
\subfloat[Cones]{\includegraphics[width=4cm]{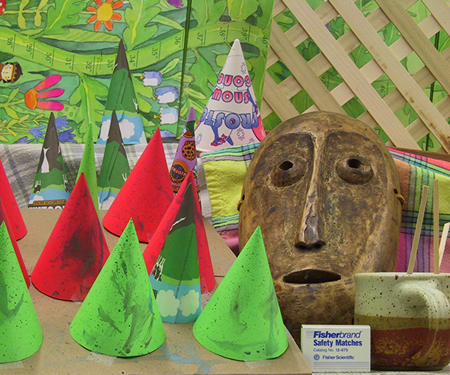}%
\label{fig_fourth_case}}
\caption{Standard Middlebury dataset \cite{bib:middleburydataset} used for evaluation of proposed algorithm}
\label{fig:test_dataset}
\end{figure}

Disparity maps for two views A and B (Fig.~\ref{fig:evalation}) have been estimated with the use of the proposed method and original unmodified DERS software. Based on views A and B and estimated disparity maps for views A and B, view V that is positioned in between of view A and B was synthesized. Exact view numbers for each of test sequences used during experiments are provided in Table \ref{table:video-views}.

The quality of estimated disparity maps for views A and B is measured as a quality of rendered view V. Quality of synthesized view V is expressed by PSNR of luminance in comparison with view V captured by real camera positioned at the same spatial position (see Fig.~\ref{fig:evalation}). 

Such methodology is compliant with experimental methodology developed and approved by the MPEG committee of International Standardization Organization and is used by other research institutes, targeted at high quality 3D television for e.g. autostereoscopic displays.

In the course of evaluation disparity maps were estimated for every frame within the sequences (mostly 250 frames per view). This has allowed to evaluate our algorithm on a wide range of different images. The disparity estimation was done with pixel, half-pixel and quarter-pixel precision. Also, a wide range of regularization terms used in Graph Cut algorithm has been evaluated. In DERS the regularization is controlled by so-called smoothing coefficient. In experiments, the range of 1 to 4 was explored.

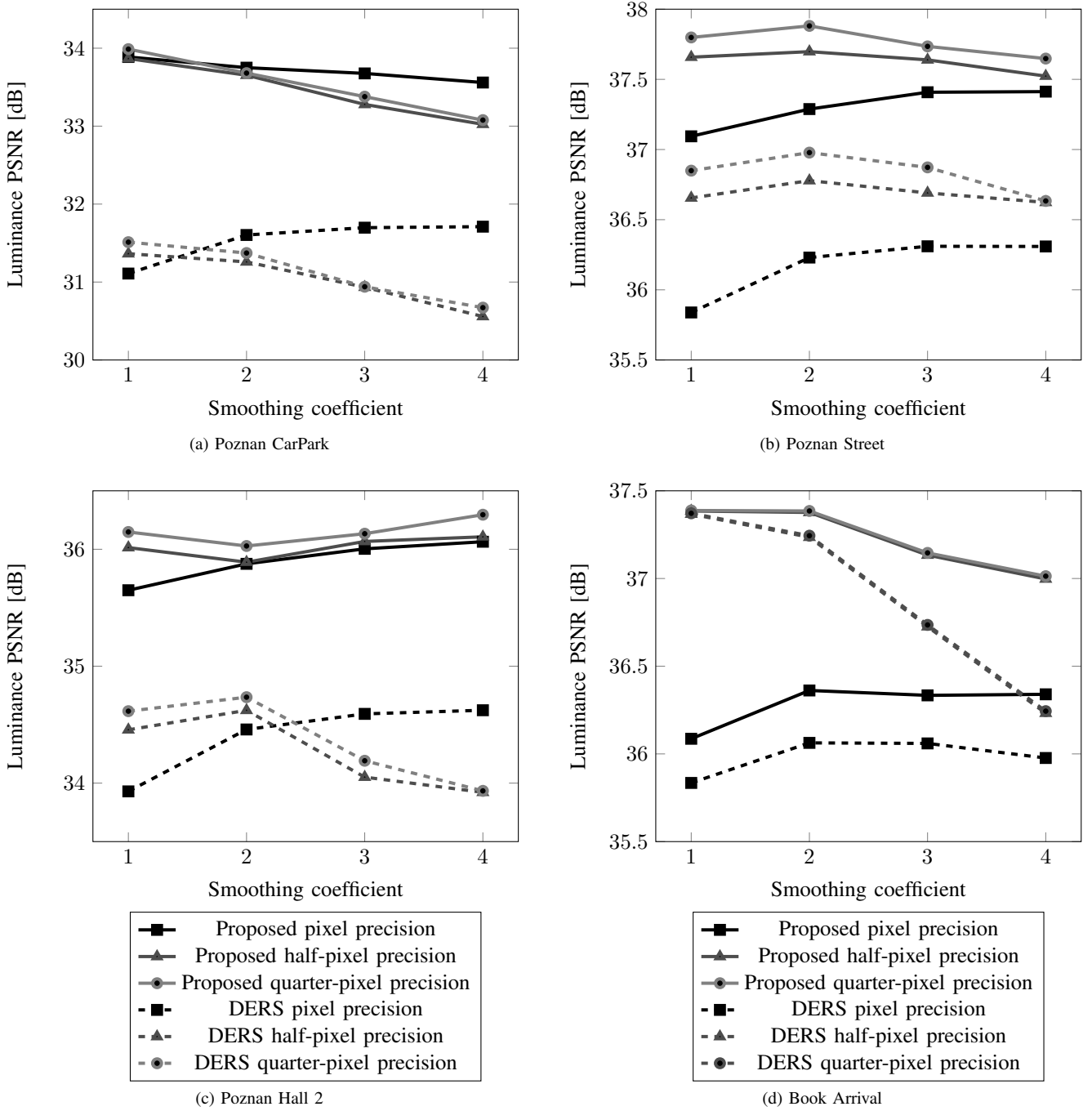
\begin{figure*}[!t]
\subfloat[Poznan CarPark]{
\begin{tikzpicture}

\pgfplotsset{
/pgfplots/area style/.style={%
area legend,
axis on top,
}}
%Poznan CarPark
\begin{axis}[
%	stack plots=y,
%    area style,
%	enlarge x limits=false,
%	enlarge y limits=false,
    enlarge x limits=0.1,
	legend style={at={(0.5,-0.2)}, anchor=north},
    xlabel={Smoothing coefficient},
    ylabel={Luminance PSNR [dB]},
    xmin=1,xmax=4,
    xtick={1,2,3,4},
    ymin=30, ymax=34.5,
    %ytick={0,10,20,30,40,50,60,70,80,90,100},
    %yticklabels={0\%,10\%,20\%,30\%,40\%,50\%,60\%,70\%,80\%,90\%,100\%},     
%   ymajorgrids=true,
]
 
\addplot [ %Proposed PEL
	draw=black!100!white,
	line width= 1.5pt,
	mark=square*,
    mark options={solid},
    ]
    coordinates {
    (1,33.887117)
    (2,33.749988)
    (3,33.676465)
    (4,33.559531)
    };

\addplot [ %Proposed HPEL
	draw=black!70!white,
	line width= 1.5pt,
	mark=triangle*,
    mark options={solid},
    ]
    coordinates {
    (1,33.86682)
    (2,33.652871)
    (3,33.278484)
    (4,33.023648)
    };
    
\addplot [ %Proposed QPEL
	draw=black!50!white,
	line width= 1.5pt,
	mark=*,
    mark options={solid},
    ]
    coordinates {
    (1,33.98781373)
    (2,33.68027529)
    (3,33.37828896)
    (4,33.07732759)
    };    

\addplot [ %DERS PEL
	draw=black!100!white,
	line width= 1.5pt,
	mark=square*,
    mark options={solid},
	dashed,
    ]
    coordinates {
    (1,31.108998)
    (2,31.60373)
    (3,31.69748)
    (4,31.710467)
    };

\addplot [ %DERS HPEL
	draw=black!70!white,
	line width= 1.5pt,
    mark=triangle*,
    mark options={solid},
    dashed,
    ]
    coordinates {
    (1,31.363611)
    (2,31.258455)
    (3,30.933475)
    (4,30.557846)
    };
    
\addplot [ %DERS QPEL
	draw=black!50!white,
	line width= 1.5pt,
    mark=*,
    mark options={solid},
    dashed,
    ]
    coordinates {
    (1,31.51165232)
    (2,31.37228093)
    (3,30.93975427)
    (4,30.67168945)
    };    

%\legend{Proposed pixel precision,Proposed half-pixel precision,Proposed quarter-pixel precision,DERS pixel precision,DERS half-pixel precision,DERS quarter-pixel precision};
%\legend{\parbox{2cm}{Proposed pixel precision},\parbox{2cm}{Proposed half-pixel precision},\parbox{2cm}{DERS pixel precision},\parbox{2cm}{DERS half-pixel precision}};  
\end{axis}
\end{tikzpicture}

\label{fig:results_poznan_carpark}}
\hfil
\subfloat[Poznan Street]{
\begin{tikzpicture}

\pgfplotsset{
/pgfplots/area style/.style={%
area legend,
axis on top,
}}
%Poznan Street
\begin{axis}[
%	stack plots=y,
%    area style,
%	enlarge x limits=false,
%	enlarge y limits=false,
    enlarge x limits=0.1,
	legend style={at={(0.5,-0.2)}, anchor=north},
    xlabel={Smoothing coefficient},
    ylabel={Luminance PSNR [dB]},
    xmin=1,xmax=4,
    xtick={1,2,3,4},
    ymin=35.5, ymax=38,
    %ytick={0,10,20,30,40,50,60,70,80,90,100},
    %yticklabels={0\%,10\%,20\%,30\%,40\%,50\%,60\%,70\%,80\%,90\%,100\%},     
%   ymajorgrids=true,
]
 
\addplot [ %Proposed PEL
	draw=black!100!white,
	line width= 1.5pt,
	mark=square*,
    mark options={solid},
    ]
    coordinates {
    (1,37.094137)
    (2,37.288223)
    (3,37.408203)
    (4,37.412559)
    };

\addplot [ %Proposed HPEL
	draw=black!70!white,
	line width= 1.5pt,
	mark=triangle*,
    mark options={solid},
    ]
    coordinates {
    (1,37.658039)
    (2,37.697465)
    (3,37.638844)
    (4,37.523215)
    };
    
\addplot [ %Proposed QPEL
	draw=black!50!white,
	line width= 1.5pt,
	mark=*,
    mark options={solid},
    ]
    coordinates {
    (1,37.79869975)
    (2,37.8810359)
    (3,37.73500924)
    (4,37.64803691)

    };    
    
\addplot [ %DERS PEL
	draw=black!100!white,
	line width= 1.5pt,
	mark=square*,
    mark options={solid},
	dashed,
    ]
    coordinates {
    (1,35.838777)
    (2,36.229984)
    (3,36.31034)
    (4,36.309176)
    };

\addplot [ %DERS HPEL
	draw=black!70!white,
	line width= 1.5pt,
    mark=triangle*,
    mark options={solid},
    dashed,
    ]
    coordinates {
    (1,36.65491)
    (2,36.778215)
    (3,36.690609)
    (4,36.623203)
    };
    
\addplot [ %DERS QPEL
	draw=black!50!white,
	line width= 1.5pt,
    mark=*,
    mark options={solid},
    dashed,
    ]
    coordinates {
    (1,36.8493552)
    (2,36.97789099)
    (3,36.87267369)
    (4,36.63346589)
    };

%\legend{Proposed pixel precision,Proposed half-pixel precision,Proposed quarter-pixel precision,DERS pixel precision,DERS half-pixel precision,DERS quarter-pixel precision};
%\legend{\parbox{2cm}{Proposed pixel precision},\parbox{2cm}{Proposed half-pixel precision},\parbox{2cm}{DERS pixel precision},\parbox{2cm}{DERS half-pixel precision}};  
\end{axis}
\end{tikzpicture}

\label{fig:results_poznan_street}}
\\
\subfloat[Poznan Hall 2]{
\begin{tikzpicture}

\pgfplotsset{
/pgfplots/area style/.style={%
area legend,
axis on top,
}}
%Poznan Hall2
\begin{axis}[
%	stack plots=y,
%    area style,
%	enlarge x limits=false,
%	enlarge y limits=false,
    enlarge x limits=0.1,
	legend style={at={(0.5,-0.2)}, anchor=north},
    xlabel={Smoothing coefficient},
    ylabel={Luminance PSNR [dB]},
    xmin=1,xmax=4,
    xtick={1,2,3,4},
    ymin=33.5, ymax=36.5,
    %ytick={0,10,20,30,40,50,60,70,80,90,100},
    %yticklabels={0\%,10\%,20\%,30\%,40\%,50\%,60\%,70\%,80\%,90\%,100\%},     
%   ymajorgrids=true,
]
 
\addplot [ %Proposed PEL
	draw=black!100!white,
	line width= 1.5pt,
	mark=square*,
    mark options={solid},
    ]
    coordinates {
    (1,35.649617)
    (2,35.875464)
    (3,36.003784)
    (4,36.064412)
    };

\addplot [ %Proposed HPEL
	draw=black!70!white,
	line width= 1.5pt,
	mark=triangle*,
    mark options={solid},
    ]
    coordinates {
    (1,36.014185)
    (2,35.889587)
    (3,36.066907)
    (4,36.10647)
    };

\addplot [ %Proposed QPEL
	draw=black!50!white,
	line width= 1.5pt,
	mark=*,
    mark options={solid},
    ]
    coordinates {
    (1,36.14774341)
    (2,36.02799437)
    (3,36.13314721)
    (4,36.29551263)
    };

\addplot [ %DERS PEL
	draw=black!100!white,
	line width= 1.5pt,
	mark=square*,
    mark options={solid},
	dashed,
    ]
    coordinates {
    (1,33.9295)
    (2,34.458887)
    (3,34.592292)
    (4,34.623374)
    };

\addplot [ %DERS HPEL
	draw=black!70!white,
	line width= 1.5pt,
    mark=triangle*,
    mark options={solid},
    dashed,
    ]
    coordinates {
    (1,34.456064)
    (2,34.622637)
    (3,34.051257)
    (4,33.92143)
    };

\addplot [ %DERS QPEL
	draw=black!50!white,
	line width= 1.5pt,
    mark=*,
    mark options={solid},
    dashed,
    ]
    coordinates {
    (1,34.61544466)
    (2,34.73502864)
    (3,34.19142653)
    (4,33.93378196)
    };    

\legend{Proposed pixel precision,Proposed half-pixel precision,Proposed quarter-pixel precision,DERS pixel precision,DERS half-pixel precision,DERS quarter-pixel precision};

%\legend{\parbox{2cm}{Proposed pixel precision},\parbox{2cm}{Proposed half-pixel precision},\parbox{2cm}{DERS pixel precision},\parbox{2cm}{DERS half-pixel precision}};  
\end{axis}
\end{tikzpicture}

\label{fig:results_poznan_hall2}}
\hfil
\subfloat[Book Arrival]{
\begin{tikzpicture}

\pgfplotsset{
/pgfplots/area style/.style={%
area legend,
axis on top,
}}
%Poznan CarPark
\begin{axis}[
%	stack plots=y,
%    area style,
%	enlarge x limits=false,
%	enlarge y limits=false,
    enlarge x limits=0.1,
	legend style={at={(0.5,-0.2)}, anchor=north},
    xlabel={Smoothing coefficient},
    ylabel={Luminance PSNR [dB]},
    xmin=1,xmax=4,
    xtick={1,2,3,4},
    ymin=35.5, ymax=37.5,
    %ytick={0,10,20,30,40,50,60,70,80,90,100},
    %yticklabels={0\%,10\%,20\%,30\%,40\%,50\%,60\%,70\%,80\%,90\%,100\%},     
%   ymajorgrids=true,
]
 
\addplot [ %Proposed PEL
	draw=black!100!white,
	line width= 1.5pt,
	mark=square*,
    mark options={solid},
    ]
    coordinates {
    (1,36.086245)
    (2,36.361646)
    (3,36.333765)
    (4,36.339751)
    };

\addplot [ %Proposed HPEL
	draw=black!70!white,
	line width= 1.5pt,
	mark=triangle*,
    mark options={solid},
    ]
    coordinates {
    (1,37.384761)
    (2,37.376621)
    (3,37.133469)
    (4,36.9972)
    };

\addplot [ %Proposed QPEL
	draw=black!50!white,
	line width= 1.5pt,
	mark=*,
    mark options={solid},
    ]
    coordinates {
    (1,37.38771804)
    (2,37.38501295)
    (3,37.14594206)
    (4,37.01324855)
    };    

\addplot [ %DERS PEL
	draw=black!100!white,
	line width= 1.5pt,
	mark=square*,
    mark options={solid},
	dashed,
    ]
    coordinates {
    (1,35.834512)
    (2,36.062871)
    (3,36.059836)
    (4,35.976633)
    };

\addplot [ %DERS HPEL
	draw=black!70!white,
	line width= 1.5pt,
    mark=triangle*,
    mark options={solid},
    dashed,
    ]
    coordinates {
    (1,37.367012)
    (2,37.235342)
    (3,36.725181)
    (4,36.231865)
    };

\addplot [ %DERS QPEL
	draw=black!70!white,
	line width= 1.5pt,
    mark=*,
    mark options={solid},
    dashed,
    ]
    coordinates {
    (1,37.37167114)
    (2,37.24412145)
    (3,36.73533996)
    (4,36.24381516)
    };

\legend{Proposed pixel precision,Proposed half-pixel precision,Proposed quarter-pixel precision,DERS pixel precision,DERS half-pixel precision,DERS quarter-pixel precision};
%\legend{\parbox{2cm}{Proposed pixel precision},\parbox{2cm}{Proposed half-pixel precision},\parbox{2cm}{DERS pixel precision},\parbox{2cm}{DERS half-pixel precision}};  
\end{axis}
\end{tikzpicture}
\label{fig:results_book_arrival}}
\caption{Performance comparison of depth estimation with use of proposed method and DERS}
\label{fig:results}
\end{figure*}

We have also evaluated our algorithm on standard Middlebury dataset \cite{bib:middleburydataset}: Tsukuba, Venus, Teddy and Cones (Fig.~\ref{fig:test_dataset}). In the course of that, we have modified DERS algorithm to directly output raw disparity maps in the format required by Middlebury evaluation webpage \cite{bib:middleweb}. Because both proposed methods and the DERS algorithm are designed to work with three input images, we have extended recommended/standard stereo pair with third image as specified in Table \ref{table:middlebury-views}.

\begin{table}[h]
\centering
\caption{Specification of three views used for disparity estimation for each Middlebury dataset.}
\label{table:middlebury-views}
\begin{tabular}{|l|c|c|c|}
\hline
\multirow{2}{*}{\textbf{Dataset name}} & \textbf{Additional view} & \multicolumn{2}{c|}{\textbf{Standard stereo pair}} \\ \cline{2-4}
        & \textbf{Left view} & \textbf{Center view} & \textbf{Right view} \\ \hline
Tsukuba & 2 & 3 & 4 \\ \hline
Venus   & 0 & 2 & 6 \\ \hline
Teddy   & 0 & 2 & 6 \\ \hline
Cones   & 0 & 2 & 6 \\ \hline
\end{tabular}
\end{table}

%\begin{figure}[!t]
%\centering
%\input{plots/poznan_carpark}
%\caption{Performance comparison of depth estimation with use of proposed method and DERS for Poznan CarPark sequence.}
%\label{fig:results_poznan_carpark}
%\end{figure}
%
%\begin{figure}[!b]
%\centering
%\input{plots/poznan_street}
%\caption{Performance comparison of depth estimation with use of proposed method and DERS for Poznan Street sequence.}
%\label{fig:results_poznan_street}
%\end{figure}
%
%\begin{figure}[!t]
%\centering
%\input{plots/poznan_hall2}
%\caption{Performance comparison of depth estimation with use of proposed method and DERS for Poznan Hall 2 sequence.}
%\label{fig:results_poznan_hall2}
%\end{figure}
%
%\begin{figure}[!b]
%\centering
%\input{plots/book_arrival}
%\caption{Performance comparison of depth estimation with use of proposed method and DERS for Book Arrival sequence.}
%\label{fig:results_book_arrival}
%\end{figure}

\section{Results}

\begin{table*}[!b]
% increase table row spacing, adjust to taste
\renewcommand{\arraystretch}{1.3}
% if using array.sty, it might be a good idea to tweak the value of
% \extrarowheight as needed to properly center the text within the cells
\caption{Quality comparison by PSNR of a synthesized view for the best depth maps with respect to smoothing coefficient.}
\label{table:results}
\centering
\begin{tabular}{|l|c|c|c|c|c|c|c|c|c|}
\hline
\multicolumn{1}{|c|}{\multirow{2}{*}{\textbf{Sequence Name}}} & \multicolumn{3}{c|}{\textbf{Pixel precison}}                                                            & \multicolumn{3}{c|}{\textbf{Half-pixel precision}}                                                      & \multicolumn{3}{c|}{\textbf{Quarter-pixel precision}}                                                   \\ \cline{2-10} 
\multicolumn{1}{|c|}{}                                        & \multicolumn{1}{c|}{\textbf{DERS {[}dB{]}}} & \multicolumn{1}{c|}{\textbf{Proposed {[}dB{]}}} & \multicolumn{1}{c|}{\textbf{Gain}} & \multicolumn{1}{c|}{\textbf{DERS {[}dB{]}}} & \multicolumn{1}{c|}{\textbf{Proposed {[}dB{]}}} & \multicolumn{1}{c|}{\textbf{Gain}} & \multicolumn{1}{c|}{\textbf{DERS {[}dB{]}}} & \multicolumn{1}{c|}{\textbf{Proposed {[}dB{]}}} & \multicolumn{1}{c|}{\textbf{Gain}} \\ \hline
Poznan Street & 36.31 & 37.41 & 1.10 & 36.78 & 37.70 & 0.92 & 36.96 & 37.88 & 0.90 \\ \hline
Poznan Hall2  & 34.62 & 36.06 & 1.44 & 34.62 & 36.11 & 1.48 & 34.74 & 36.39 & 1.56 \\ \hline
Poznan CarPark& 31.71 & 33.89 & 2.18 & 31.36 & 33.87 & 2.50 & 31.51 & 33.99 & 2.48 \\ \hline
Book Arrival  & 36.06 & 36.36 & 0.30 & 37.37 & 37.38 & 0.02 & 37.37 & 37.39 & 0.02 \\ \hline
\textbf{Average} & \multicolumn{2}{c|}{-} & \textbf{1.26} & \multicolumn{2}{c|}{-} & \textbf{1.23} & \multicolumn{2}{c|}{-} & \textbf{1.24} \\ \hline
\end{tabular}
\end{table*}

The comparison of quality of estimated disparity maps for proposed method versus original DERS can be found in Fig.~\ref{fig:results_book_arrival}, \ref{fig:results_poznan_carpark}, \ref{fig:results_poznan_hall2}, \ref{fig:results_poznan_street}. As it can be noticed, the smoothing coefficient can have significant impact on the quality of disparity maps estimated by DERS. It can be expected that in a real-world-use scenario, this parameter will be automatically controlled to provide the best results. Therefore, in summarized Table \ref{table:results}, we have presented only the best-performing cases. Depending on the case, the proposed occlusion handling brings a gain of 0.02-2.50 dB of luminance PSNR of synthesized view, related to the original unmodified DERS. On average, the proposal provides an improvement of 1.26 dB for pixel-precise disparity estimation, 1.23 dB for half-precise disparity estimation, and 1.18 db for quarter-precise disparity estimation.

%\begin{figure}[!h]
%\centering
%\input{plots/gains}
%\caption{Depth maps quality improvement estimated with use of proposed method over DERS.}
%\label{fig:results_gains}
%\end{figure}

The application of proposed occlusion handing to Middlebury images results in 0.2 bad pixel improvement (Table~\ref{table:results2}). Please keep in mind that Middleburry datasets have very little occlusions. 

\begin{table*}[!b]
% increase table row spacing, adjust to taste
\renewcommand{\arraystretch}{1.3}
% if using array.sty, it might be a good idea to tweak the value of
% \extrarowheight as needed to properly center the text within the cells
\caption{Comparison of proposed method with DERS on middlebury datasets.}
\label{table:results2}
\centering
\begin{tabular}{|l|lll|lll|lll|lll|}
\hline
\multicolumn{1}{|c|}{\multirow{2}{*}{Algorithm}} & \multicolumn{3}{c|}{Tsukuba} & \multicolumn{3}{c|}{Venus} & \multicolumn{3}{c|}{Teddy} & \multicolumn{3}{c|}{Cones}  \\ \cline{2-13} 
\multicolumn{1}{|c|}{} & \multicolumn{1}{c}{nonocc} & \multicolumn{1}{c}{all} & \multicolumn{1}{c|}{disc} & \multicolumn{1}{c}{nonocc} & \multicolumn{1}{c}{all} & \multicolumn{1}{c|}{disc} & \multicolumn{1}{c}{nonocc} & \multicolumn{1}{c}{all} & \multicolumn{1}{c|}{disc} & \multicolumn{1}{c}{nonocc} & \multicolumn{1}{c}{all} & \multicolumn{1}{c|}{disc} \\ \hline
GC+occ & 1.19 & 2.01 & 6.24 & 1.64 & 2.19 & 6.75 & 11.2 & 17.4 & 19.8 & 5.36 & 12.4 & 13.0 \\ \hline
2OP+occ [36] & 2.91 & 3.56 & 7.33 & 0.24 & 0.49 & 2.76 & 10.9 & 15.4 & 20.6 & 5.42 & 10.8 & 12.5 \\ \hline 
Putv3 & 1.77 & 3.86 & 9.42 & 0.42 & 0.95 & 5.72 & 7.02 & 14.2 & 18.3 & 2.40 & 9.11 & 6.56 \\ \hline
CostAggr+occ [37]  & 1.38 & 1.96 & 7.14 & 0.44 & 1.13 & 4.87 & 6.80 & 11.9 & 17.3 & 3.60 & 8.57 & 9.36 \\ \hline  
DERS &  2.70 & 3.30 & 12.10 & 0.67 & 1.25 & 8.53 &  10.2 & 11.5 & 23.3 & 5.17 & 7.33 &  9.50 \\ \hline
Proposed & 2.65 & 3.01  & 11.20 & 0.63 & 1.02 & 8.34 & 9.96 & 10.97 & -.-- & 5.02 & 7.12 & -.-- \\ \hline
\end{tabular}
\end{table*}

\section{Conclusion}
We have presented a novel approach to occlusion handling in disparity estimation, based on a modification of similarity cost function. Proposed approach has been tested in the three-view disparity estimation scenario. For occlusion detection synthesized disparity maps of left and right views have been used. 

For well-known multiview video test sequences, the experimental results show that the proposed approach provides virtual view quality improvement of 1.25 dB of luminance PSNR over the state-of-the-art technique implemented in MPEG Depth Estimation Reference Software (DERS). Moreover, direct quality evaluation of estimated disparity reveals that proposed the approach reduces a number of bad pixels by 1.26 p.p. 

% use section* for acknowledgement
\section*{Acknowledgment}

Research project was supported by National Science Centre, Poland, according to the decision DEC-2012/05/N/ST7/1279.

% Can use something like this to put references on a page
% by themselves when using endfloat and the captionsoff option.
\ifCLASSOPTIONcaptionsoff
  \newpage
\fi


\begin{thebibliography}{1}

\bibitem{bib:poznan_sequences}
M.~Doma\'{n}ski, O.~Stankiewicz, K.~Wegner et~al., \emph{Poznań multiview video test sequences and camera parameters}, ISO/IEC JTC1/SC29/WG11 Doc. M17050, Xian, China, October 2009.

\bibitem{bib:hhi_sequences}  
I.~Feldmann, A.~Smolic, et~al., \emph{HHI Test Material for 3D Video}, ISO/IEC JTC1/SC29/WG11, Doc. M15413, Archamps, France, 2008.

\bibitem{bib:depth_similarity}  
K.~Wegner, O.~Stankiewicz, \emph{Similiarity measures for depth estimation}, 3DTV-Conference 2009, Potsdam, Germany, May 2009.

\bibitem{bib:depth_similarity2}  
S.~Birchfield and C.~Tomasi, \emph{A pixel dissimilarity measure that is insensitive to image sampling}, IEEE Transactions on Pattern Analysis and Machine Intelligence, 20(4):401–406, April 1998.

\bibitem{bib:occ1} 
Woo-Seok Jang, Yo-Sung Ho, \emph{Efficient disparity map estimation using occlusion handling for various 3D multimedia applications}, IEEE Consumer Electronics, vol. 57, no. 4, pp.1937,1943, November 2011.

\bibitem{bib:mvd} 
K.~Müller, P.~Merkle, T.~Wiegand, \emph{3-D video representation using depth maps}, Proceedings of the IEEE, vol. 99, no. 4, April 2011.

\bibitem{bib:dibr}
L.~Zhang and W.~J.~Tam, \emph{Stereoscopic image generation based on depth images for 3DTV}, IEEE Trans. on Broadcasting, vol. 51, no. 2, pp. 191-199, June 2005.

\bibitem{bib:tof}
S.~Y.~Kim, J.~H.~Cho, and A. ~oschan, \emph{3D video generation and service based on a TOF depth sensor in MPEG-4 multimedia framework}, IEEE Trans. on Consumer Electronics, vol. 56, no. 3, pp. 1730-1738, August 2010.

\bibitem{bib:naszftv}
M.~Doma\'{n}ski, A.~Dziembowski, A.~Kuehn, M.~Kurc, A.~\L uczak, D.~Mieloch, J.~Siast, O.~Stankiewicz and K.~Wegner, \emph{Experiments on acquisition and processing of video for free-viewpoint television}, 3DTV Conference 2014, Budapest, Hungary, July 2014.

\bibitem{bib:hartley}
R.~Hartley and A.~Zisserman, \emph{Multiple View Geometry in Computer
Vision}, 2nd ed., Cambridge University Press, 2003, pp. 262-278.

\bibitem{bib:muliviewmatching1}
M.~Okutomi and T.~Kanade. \emph{A multiple baseline stereo},
IEEE Trans. PAMI, 15(4):353–363, April 1993.

\bibitem{bib:muliviewmatching2}
R.~T.~Collins. \emph{A space-sweep approach to true multi-image
matching}, In CVPR’96, pp. 358–363, San Francisco, 1996.

\bibitem{bib:rectify1}
J.~Stankowski, K.~Klimaszewski, O.~Stankiewicz, K.~Wegner, M.~Doma\'{n}ski, \emph{Preprocessing methods used for Poznan 3D/FTV test sequences}, ISO/IEC JTC1/SC29/WG11 MPEG 2010/M17174, Doc. m17174, Kyoto, Japan, January 2010.

\bibitem{bib:rectify2}
J.~Stankowski, K.~Klimaszewski, Book: "Image Processing and Communications Challenges 2", Chapter: "Application of epipolar rectification algorithm in 3D Television", Advances in Intelligent and Soft Computing: Vol. 84, Springer-Verlag, Berlin, 2010, pp. 345-352, ISBN: 978-3-642-16294-7. 

\bibitem{bib:dersold}
M.~Tanimoto, T.~Fujii, K.~Suzuki, N.~Fukushima, and Y.~Mori, \emph{Reference softwares for depth estimation and view synthesis}, ISO/IEC JTC1/SC29/WG11, Doc. M15377, Archamps, France, April 2008.

\bibitem{bib:de3view-ho}
Sang-Beom Lee, Yo-Sung Ho, \emph{Multi-view Depth Map Estimation
Enhancing Temporal Consistency}, 23rd International Technical
Conference on Circuits/Systems, Computers and Communications.

\bibitem{bib:de3view-tanimoto}
M.~Wildeboer, N.~Fukushima, T.~Yendo, M.~T.~Panahpour, T.~Fujii, M.~Tanimoto, \emph{A semi-automatic multi-view depth estimation method}, Proceedings of the SPIE, Volume 7744, 2010.  

\bibitem{bib:ders}
M.~Wildeboer, O.~Stankiewicz, K.~Wegner, \emph{A soft-segmentation matching in Depth Estimation Reference Software (DERS) 5.0}, ISO/IEC JTC1/SC29/WG11 Doc. M17049, Xian, China, October 2009.

\bibitem{bib:bp1}
J.~Sun, N.~N.~Zheng, and H.~Y~Shum, \emph{Stereo matching using belief
propagation}, IEEE Trans. on Pattern Analysis and Machine
Intelligence, vol. 25, no. 7, pp. 787-800, July 2003.

\bibitem{bib:bp2}
P.~Felzenszwalb and D.~Huttenlocher, \emph{Efficient belief propagation for early vision}, In Proc. IEEE Computer Society Conference on Computer
Vision and Pattern Recognition, pp.261-268, June 2004.

\bibitem{bib:gc}
Y.~Boykov, O.~Veksler, and R.~Zabih, \emph{Fast approximate energy
minimization via graph cuts}, IEEE Trans. on Pattern Analysis and
Machine Intelligence, vol. 23, no. 11, pp.1222-1239, November 2001.

\bibitem{bib:dp}
O.~Veksler, \emph{Stereo correspondence by dynamic programming on a tree} CVPR 2005.

\bibitem{bib:crosscheck}
G.~Egnal and R.~Wildes, \emph{Detecting binocular halfocclusions: empirical comparisons of five approaches}, IEEE Trans. on Pattern Analysis and Machine Intelligence, vol. 24 no. 8, pp. 1127-1133, August 2002.

\bibitem{bib:ordering}
A.~Bobick and S.~Intille, \emph{Large occlusion stereo}, Int. Journal of Computer Vision, vol.33, no. 3, pp. 181-200, September 1999.

\bibitem{bib:uniqness}
D.~Marr and T.~A.~Poggio, \emph{Cooperative computation of stereo disparity}, Science, vol. 194, no. 4262, pp. 283-287, October 1976.

\bibitem{bib:uniqness2}
V.~Kolmogorov and R.~Zabih, \emph{Computing visual correspondence with occlusions using graph cuts}, In Proc. IEEE International Conference on Computer Vision, pp. 508-515, July 2001.

\bibitem{bib:occ1a}
T.~Liu, P.~Zhang, and L.~Luo, \emph{Dense stereo correspondence with contrast context histogram, segmentation-based two-pass aggregation and occlusion handling}, Lecture Notes in Computer Science, vol. 5414,pp. 449-461, January 2009. 

\bibitem{bib:occ2}
R.~Ben-Ari and N.~Sochen, \emph{Stereo matching with Mumford-shah regularization and occlusion handling}, IEEE Trans. on Pattern Analysis and Machine Intelligence, vol. 32, no. 11, pp. 2071-2084, November 2010.

\bibitem{bib:middleburydataset}
D.~Scharstein and R.~ Szeliski, \emph{High-Accuracy Stereo Depth Maps Using Structured Light}, In IEEE Computer Society Conference on Computer Vision and Pattern Recognition (CVPR 2003), volume 1, pages 195-202, Madison, WI, June 2003. 

\bibitem{bib:middleweb}
Middlebury Stereo Evaluation - Version 2, webpage visited 2015-01-24, http://vision.middlebury.edu/stereo/eval .

\bibitem{bib:3dhevc}
G.~Tech, K.~Wegner, Y.~Chen, S.~Yea, \emph{3D-HEVC Draft Text 6} Joint Collaborative Team on 3D Video Coding Extension Development of ITU-T SG 16 WP 3 and ISO/IEC JTC 1/SC 29/WG 11 Doc. JCT3V-J1001, 10th Meeting: Strasbourg, FR, 18–24 October 2014.

\bibitem{bib:mvc}
Annex I \emph{Multiview and Depth video coding} of ISO/IEC 14496-10, Int. Standard \emph{Generic coding of audio-visual objects – Part 10: Advanced Video Coding}, 8th Ed., 2013, also: ITU-T Rec. H.264, Edition 8.0, 2013.

\bibitem{bib:3davc}
\emph{3D-AVC Draft Text 9}, JCT-3V of ITU T SG 16 WP 3 and ISO/IEC JTC 1/SC 29/WG 11, Doc. JCT3V-G1003, San Jose, USA, 2014.

\bibitem{bib:steromatching}
D.~Scharstein and R.~Szeliski, \emph{A taxonomy and evaluation of dense two-frame stereo correspondence algorithms}, International Journal of Computer Vision, 47(1/2/3):7-42, April-June 2002.

\bibitem{bib:multiviewmatching}
S. M. Seitz, B. Curless, J. Diebel, D. Scharstein, and R. Szeliski, \emph{A comparison and evaluation of multi-view stereo reconstruction
algorithms}, in Proceedings of the IEEE Computer Society Conference on Computer Vision and Pattern Recognition (CVPR’06), pp. 519–526, June 2006.

\bibitem{bib:os}
O.~Stankiewicz, "Stereoscopic depth map estimation and coding techniques for multiview video systems", PhD Dissertation at Poznan University of Technology, Faculty of Electronics and Telecommunications, 2014. 



%http://vclab.gist.ac.kr/papers/01/2014/10.1007_s11760-013-0550-2.pdf

\end{thebibliography}
\end{document}